\input epsf
\documentstyle[preprint,aps,pre]{revtex}
\tighten
\begin{document}
\draft

\title{Universal Short-Time Behavior in
Critical Dynamics\\[2mm]
near Surfaces}
\author{U. Ritschel and P. Czerner}

\address{Fachbereich Physik, Universit\"at GH Essen, 45117 Essen
(F\ R\ Germany)}

\maketitle
\vspace*{4cm}
\narrowtext

\begin{abstract}
We study the time evolution of classical spin systems
with purely relaxational dynamics, quenched from
$T\gg T_c$ to the critical point,
in the semi-infinite geometry. Shortly after the quench,
like in the bulk, a nonequilibrium regime governed
by universal power laws
is also found near the surface.
We show for `ordinary' and `special' transitions
that the corresponding critical exponents
differ from their bulk values, but can be expressed
via scaling relations in terms of known
bulk and surface exponents.
To corroborate our scaling analysis, we present
perturbative ($\epsilon$-expansion) and Monte Carlo results.
\end{abstract}
The dynamics of spin systems after quenches from
a high-temperature initial state to or below
a critical point has attracted a great deal of interest during recent
years \cite{bray}.
In particular, for
quenches from $T\gg T_c$
to the critical temperature Janssen et al. \cite{jans}
found that
in Ising-like systems with purely relaxational dynamics
(model A) \cite{halp} a previously unknown regime
shortly after the quench with
unexpected features exists.
In its most pronounced form the so-called universal short-time behavior
(USTB) appears in the time dependence of the order parameter itself.
Suppose the initial state carries some small magnetization
$m_0$. Then, as Janssen et al.\cite{jans} showed, after the
quench the magnetization will grow for a {\it macroscopic} time span,
governed by the universal power law
\begin{equation}\label{ustb}
m(t)\sim m_0\,t^{\theta}$\qquad\mbox{with}\quad
$\theta=(x_0-\beta/\nu)/\zeta\>,
\end{equation}
where $x_0$ is the scaling dimension of the {\it initial} field
\cite{foo1,own} (whereas the
static exponents have their usual meaning, and
$\zeta$ denotes the dynamic exponent).
Thus, technically spoken, USTB is caused
by the influence of the scaling behavior of the initial field on
(in macroscopic units) early stages of the relaxational process\cite{own}.
The corresponding operator dimension $x_0$ is a genuine new exponent.
Especially for the $n$-vector
model in $2<d<4$
it turns out that $x_0>\beta/\nu$, and thus the magnetization
really increases initially, up to a characteristic time
$t_0\sim m_0^{-\zeta/x_0}$. After that it
crosses over to the well-known
algebraic decay $m\sim t^{-\beta/\nu \zeta}$\cite{jans,own2}.
Recently, the scenario developed in Ref.\,\cite{jans} was
directly confirmed by means of Monte Carlo simulation \cite{monte},
and, based on USTB, a procedure to determine static
critical exponents was suggested \cite{liea}.

Another quantity where the short-time exponent
$\theta$ enters is the
autocorrelation function $A(t)=\langle \phi(t)\,\phi(0)\rangle
\sim t^{-\lambda/\zeta}$. It was shown by Janssen \cite{jans2} that $\lambda
=d-\zeta\theta$,
and it is this quantity where USTB
was actually
first seen in a numerical simulation\cite{huse}.

Similar in many respects to initial conditions for time-dependent
processes are boundary conditions at surfaces.
It is well known that critical behavior is modified
near surfaces; each bulk universality class
splits up in three surface universality classes, depending
on whether at bulk criticality the surface is disordered,
ordered, or also critical. The corresponding universality classes
are called ordinary, extraordinary, and
special transition, respectively \cite{diehl}.
In this Letter we focus attention on the influence of
`ordinary' and `special'
surfaces on the time-dependence of $m$ and $A$.
We show that close to the surface the USTB
is modified depending on the surface universality class.
Further, we show that the values of the
`new' exponents that govern the short-time dynamics near the surface
can be expressed in terms of scaling relations
between other known critical exponents.
In a recent preprint, Majumdar and Sengupta\cite{maju}
correctly determined the surface behavior of the
autocorrelation function
for the $n$-vector model, but the underlying
scaling relation was not recognized.

In the framework of field theory the
modification of critical behavior near surfaces
is the result of additional
{\it primitive}
uv-divergences, which are located in the surface \cite{diehl}.
These divergences need
additional $Z$-factors, which, in turn, give rise to independent
surface exponents. As shown by Janssen et al. \cite{jans}, the situation
is largely analogous at the time `surface' $t=0$.
That in general $x_0\neq\beta/\nu$ is the result of an
additional {\it logarithmic} divergence at $t=0$
in the upper critical dimension $d^*=4$. What happens,
if we have both initial and boundary conditions?
It is well known that the degree of surface divergences
is reduced at least by one compared with
bulk divergences\cite{diehl}. Since the bulk
divergence at $t=0$ is just logarithmic,
it is immediately clear that
no additional primitive divergence located at $z=t=0$ exists, and, thus,
the USTB near surfaces must be governed
by known bulk, bulk short-time, and surface exponents.

To be more specific,
let us consider the $t$- and $z$-dependent magnetization in
the semi-infinite system. At the bulk
critical point it follows from dimensional analysis and
renormalization-group (RG) invariance that asymptotically
\begin{equation}\label{scal}
m(z,t,m_0)\sim t^{-\beta/\nu\zeta}\,{\cal F}(z/t^{1/\zeta},m_0\,
t^{x_0/\zeta})\>.
\end{equation}
As usual all quantities are made dimensionless
with the help of an appropriate power of the renormalization mass
$\mu$, and we set $\mu=1$ afterwards.
Because of different scaling dimensions,
$m_0$ is {\it not} the initial value of $m$. Instead, as
close as possible to the naive expectation, one finds that
$m\sim m_0$ for $t\ll t_0$\cite{foo2}.
It follows that
${\cal F}(x, y)\sim y$ for
$x\rightarrow \infty$ and $y\rightarrow 0$, and we obtain the
known bulk short-time behavior of Eqn. (\ref{ustb}).

When do we expect modifications due to the surface?
The surface will influence the dynamics of a spin
when
the nonequilibrium (growing)
ation length $\xi(t)\sim t^{1/\zeta}$
has become larger than the spin's distance $z$ from the surface.
This is analogous to the static
situation, where
surface critical behavior is observed if the
equilibrium correlation length is larger than z\cite{diehl}.
Now, if $z\ll \xi(t)$, the short-distance expansion (SDE) \cite{sde,diehl}
\begin{equation}\label{sdem}
m(z,t,m_0) \sim z^{(\beta_1-\beta)/\nu}\,m_1(t,m_0)
\end{equation}
should hold, where $\beta_1/\nu$
is the scaling dimension
of the surface field \cite{foox} and $m_0$ {homogeneous}.
Moreover, exploiting as before the RG-invariance
leads to
$m_1(t,m_0)\sim t^{-\beta_1/\nu z} {\cal F}_1(m_0\, t^{x_0/z})$.
Demanding again that $m\sim m_0$ for $t\ll t_0$,
it follows that ${\cal F}_1(y) \sim y$ for $y\rightarrow 0$.
As a result, the magnetization close to the surface ($z\ll\xi(t)$)
takes the form
\begin{mathletters}\label{first}
\begin{equation}\label{magn}
m(z,t,m_0)\sim m_0\,z^{(\beta_{1}-\beta)/\nu}\,t^{\theta_1}\, ,
\end{equation}
with
\begin{equation}\label{scalrel1}
\theta_1=(x_0-\beta_1/\nu)/\zeta= \theta+(\beta-\beta_1)/\nu \zeta\>.
\end{equation}
\end{mathletters}
In other words, the short-time exponent near the surface is
determined by the difference between the scaling dimensions
of the initial field and the surface field. Eqn. (\ref{magn})
holds if $t\ll t_0$ {\it and} $z\ll \xi(t)$. If, on the other hand,
$z \ge \xi(t_0)$
only bulk short-time behavior will be seen and eventually a
crossover to the nonlinear surface behavior $m\sim t^{-\beta_1/\nu \zeta}$
\cite{didi,kiku}.

Next we set $m_0=0$ and
consider the autocorrelation function $A(z,t):=
\langle\phi({\bf r},t)\phi({\bf r},0)\rangle$.
Because of the Dirichlet initial condition the asymptotic result for
$A$ vanishes identically, and,
as shown by Janssen \cite{jans2}, one has to consider
the leading correction in an expansion in the inverse
of the initial temperature $\tau_0^{-1}$:
\begin{equation}\label{auto}
A(z,t)= \tau_0^{-1}\langle \phi({\bf r},t)\, \tilde \phi_0({\bf r})\rangle\>.
\end{equation}
$\tilde \phi_0$ is the initial response field with
bulk scaling dimension $d-x_0$ \cite{own}.

For $z\ll \xi(t)$ both operators in (\ref{auto})
can be replaced by their SDE. The SDE of the order-parameter field
can be read from Eqn.\ (\ref{sdem}).
The SDE of $\tilde \phi_0$ is obtained as follows:
It is known from the equilibrium dynamics \cite{didi,diehl} that
the response field $\tilde \phi$ needs an additional
renormalization factor $Z_1^{1/2}$ at the surface, the same
$Z$-factor that is required to renormalize the order-parameter field
at the surface.
Utilizing our earlier result that there are no new
primitive divergencies at $z=t=0$,
we conclude that an additional $Z_1^{1/2}$ also suffices to renormalize
$\tilde \phi_0$ at the surface.
As a consequence, $\tilde \phi_0$
has the same SDE as $\phi$, and
the autocorrelation function takes the form
\begin{mathletters}\label{second}
\begin{equation}
A(z,t)\sim z^{2(\beta_1-\beta)/\nu}\; t^{-\lambda_1/\zeta}\, ,
\end{equation}
with
\begin{equation}\label{scalrel2}
\lambda_1=d-\zeta\theta+2(\beta_1-\beta)/\nu \>.
\end{equation}
\end{mathletters}

Eqs. (\ref{first}) and (\ref{second}) are the essential results of this
Letter.
Numerical values for $\theta_1$ and $\lambda_1$, where
we have used the literature values of the exponents involved, are
given in Table 1 (together with other results discussed below).

In order to confirm Eqs. (\ref{first}) and (\ref{second}), we
present a first-order $\epsilon$-expansion (one-loop)
for the equation of motion
of the $n$-vector model A \cite{halp}.
According to the above discussion, $m$ is
small for $t\ll t_0$, and so we may consider a linear
approximation (in $m$).
It takes the form
\begin{equation}\label{eost}
\dot m+\Gamma\tau\,m-\Gamma \,m''+
\frac{\Gamma g}{6} (n+2) C(z,t)\,m =0\>,
\end{equation}
where C(z,t) is the one-loop (tadpole) contribution.
According to the standard procedure \cite{diehl}
$C(z,t)$ can be separated
in a bulk and a surface part, where the latter
is determined by the bulk propagator at the image point:
\begin{mathletters}
\begin{equation}\label{oneloop}
C(z,t)=C_b(0,t)\pm C_b(2z,t)\,.
\end{equation}
In this equation `$+$' stands for special and the `$-$' for ordinary
transition, and
\begin{equation}
C_b(z,t)=\frac{2\Gamma}{(2\pi)^d}
\int d^{d}p\>{\rm e}^{-2\Gamma({\bf p}_{\parallel}^2
+p_{\perp}^2+\tau_0)(t-t')+{\rm i}2p_{\perp}z}\,.
\end{equation}
\end{mathletters}
The first term in (\ref{oneloop}), the bulk contribution,
contains the usual
uv-divergence. Without going into technical details, we state
that this divergence (after dimensional regularization) can be absorbed
in the bare temperature by minimal subtraction as in the static calculation.
The second term
on the right-hand side of (\ref{oneloop}) is the surface contribution,
which is finite for any $z>0$. So after renormalization
we obtain a finite equation, and
at the RG fixed point $g^*=48 \pi^2  \epsilon/(n+8)$ and $\tau=0$
it takes the form
\begin{equation}\label{fipo}
\dot m -m'' + \theta \left(-\frac{1}{\tilde t}
\pm \frac{1}{z^2} {\rm e}^{-z^2/\tilde t}\right)\,m=0,\quad
\end{equation}
where
$\theta=\epsilon\,(n+2)/(4n+32)$ is the one-loop
result for the bulk short-time exponent\cite{own,jans2}
and with $\tilde t=2\Gamma t$.

It is straightforward to show that for uniform initial magnetization
$m_0$  Eqn.\,(\ref{fipo})
has solutions of the form $m(\tilde t,z)=
m_0\,U(\tilde t)\,V(z^2/\tilde t)$.
Let us first consider the bulk limit $z\gg \xi(\tilde t)$ with
$\xi\sim \tilde t^{1/2}$ ($\zeta=2$ in one-loop).
By straightforward analysis it follows that $U\sim \tilde t^{\theta}$
and $V\rightarrow\mbox{const.}$, and thus $m\sim t^{\theta}$. This
is the bulk short-time behavior.

What happens if we move closer to the surface?
For $z\ll \xi(\tilde t)$ the function $V(x)$ behaves as
a power of $x$, which now depends
on the universality class of the surface.
For the ordinary transition one finds
$V\sim x^{1/2-\theta}$. This means that
$m\sim m_0\,z^{1-2\theta}\,\tilde t^{-1/2+2\theta}$
and $\theta_1^{ord}=-1/2+\epsilon\,(n+2)/(2n+16)$.
For the special transition the result
is $V\sim x^{-\theta}$ and
$m\sim m_0\,z^{-2\theta}\,\tilde t^{2\theta}$ such that
$\theta_1^{sp}=\epsilon\,(n+2)/(2n+16)$. The one-loop values of
$\theta_1^{ord}$ and $\theta_1^{sp}$
are consistent with (\ref{scalrel1}).
That in this simple approximation the surface exponents can effectively
be expressed with the help of the bulk
$\theta$ is an artifact of the
one-loop approximation.

We do not repeat the calculation of Ref.\,\cite{maju}
for the
autocorrelation function here. The one-loop results
$\lambda_1^{ord}=6-\epsilon\,(5n+22)/(2n+16)$
and $\lambda_1^{sp}=4-\epsilon\,(5n+22)/(2n+16)$
derived in \cite{maju} are both consistent with
the scaling relation (\ref{scalrel2}).
The figures for $n=\epsilon=1$ are
given in the Table.

{}From the one-loop results we can immediately read off the {\it exact}
exponents for $n\rightarrow \infty$. They are given by
$\theta_1^{ord}=(3-d)/2$, $\theta_1^{sp}=(4-d)/2$,
$\lambda_1^{ord}=(5d-8)/2$, and $\lambda_1^{sp}=(5d-12)/2$.
Numerical values for $d=3$ are displayed in the Table.

To summarize these results, at the `special' surface in $2<d<4$
the order grows always faster than in the bulk, and the autocorrelations
decay slower. For `ordinary' surfaces,
the USTB for the Ising universality class
is drastically different from bulk behavior; the order starts
to decay from the beginning. On the other hand,
for $n\rightarrow \infty$ we find  $\theta_1^{ord}>0$ for $d<3$.
Autocorrelations generally decay faster at `ordinary' surfaces.\\[5mm]

\hspace*{1.8cm}\vspace*{2mm}
\begin{tabular}{||p{1.5cm}||p{2.0cm}||p{1.5cm}|p{1.9cm}|p{1.7cm}||}\hline
& bulk &surface & $\theta_1$ & $\lambda_1/\zeta$\\\hline\hline
1-loop& $\theta=0.08$ & {\it ord}& -0.33 &2.25\\\cline{3-5}
($n=1$)       & $\lambda/\zeta=1.42$ & {\it sp} & 0.17 & 1.25\\\hline
MC   &    & {\it ord}& -0.274(13)&1.94(16)\\\cline{3-5}
(Ising)     &  & {\it sp} & 0.171(7)&1.12(7)\\\hline
Lit.   & $0.104$ & {\it ord}& -0.255 & 1.732 \\\cline{3-5}
(Ising)       & $1.372$ &{\it sp} & 0.176&1.23\\\hline
$n\rightarrow \infty$ &$1/4$  & {\it ord}& 0 & 7/2 \\\cline{3-5}
(exact)     &$ 5/2 $  & {\it sp} & 1/2 & 3/2\\\hline
\end{tabular}\\[4mm]

\noindent
{\small {\bf Table 1:} Dynamic surface exponents $\theta_1$ and
$\lambda_1/\zeta$
for the d=3 Ising model ($n=1$) for `ordinary' and `special' surfaces.
For comparison some bulk results are also displayed.
The one-loop approximation has been extrapolated to $\epsilon=1$.
In section `MC' our Monte Carlo estimates are displayed.
In `Lit.' results obtained with (\ref{scalrel1})
and (\ref{scalrel2}) in combination with the
best known literature values
for the exponents involved are shown. The latter are taken from
Refs.\,\cite{diehl,grass,ruge}.
In the last section the exact exponents of the $n\rightarrow
\infty$ limit
in $d=3$ are displayed.}\\[3mm]

Finally, we present a Monte Carlo simulation of the
$d=3$ Ising model. We have used
the heat-bath algorithm for nonconserved spins
and the usual time-dependent
interpretation of our data \cite{bind}.
The simulations were carried out
on a lattice of size $20^2\times 40$, where we have implemented
the nontrivial boundary conditions on the surfaces perpendicular
to the long axis and periodic boundary conditions in the remaining
directions. The high-temperature initial configuration was generated
as in Ref.\,\cite{monte} by randomly selecting and
flipping $1/2-m_0/2$ of the
spins in an initially completely (in $+$-direction)
ordered configuration. To determine the exponent $\theta$ from the
order-parameter profiles, we started from an initial state with
nonzero $m_0$; for the data displayed in Fig.\,1 we
chose $m_0=0.04$. For system size tending to infinity,
$m_0$ would scale to zero \cite{monte}.
For $A(t)$ a disordered initial state (with $m_0=0$) was prepared.
All simulations
have been carried out at the bulk critical value $J/k_BT_c=0.2216$
\cite{ruge}.
The Monte Carlo results for $\theta_1$ and $\lambda_1$ were
obtained by averaging over 200,000 independent histories.
To estimate the errors of the exponents, we have divided the
data into 20 runs.
To simulate the special transition, the result
of Ruge et al. \cite{ruge} for the critical surface coupling
$J_{1}/J=1.5004$ was adopted.
In the case of the ordinary transition we chose a value
somewhat lower than the bulk coupling\cite{foo3}.\\[3mm]

\def\epsfsize#1#2{0.58#1}
\hspace*{0.8cm}\epsfbox{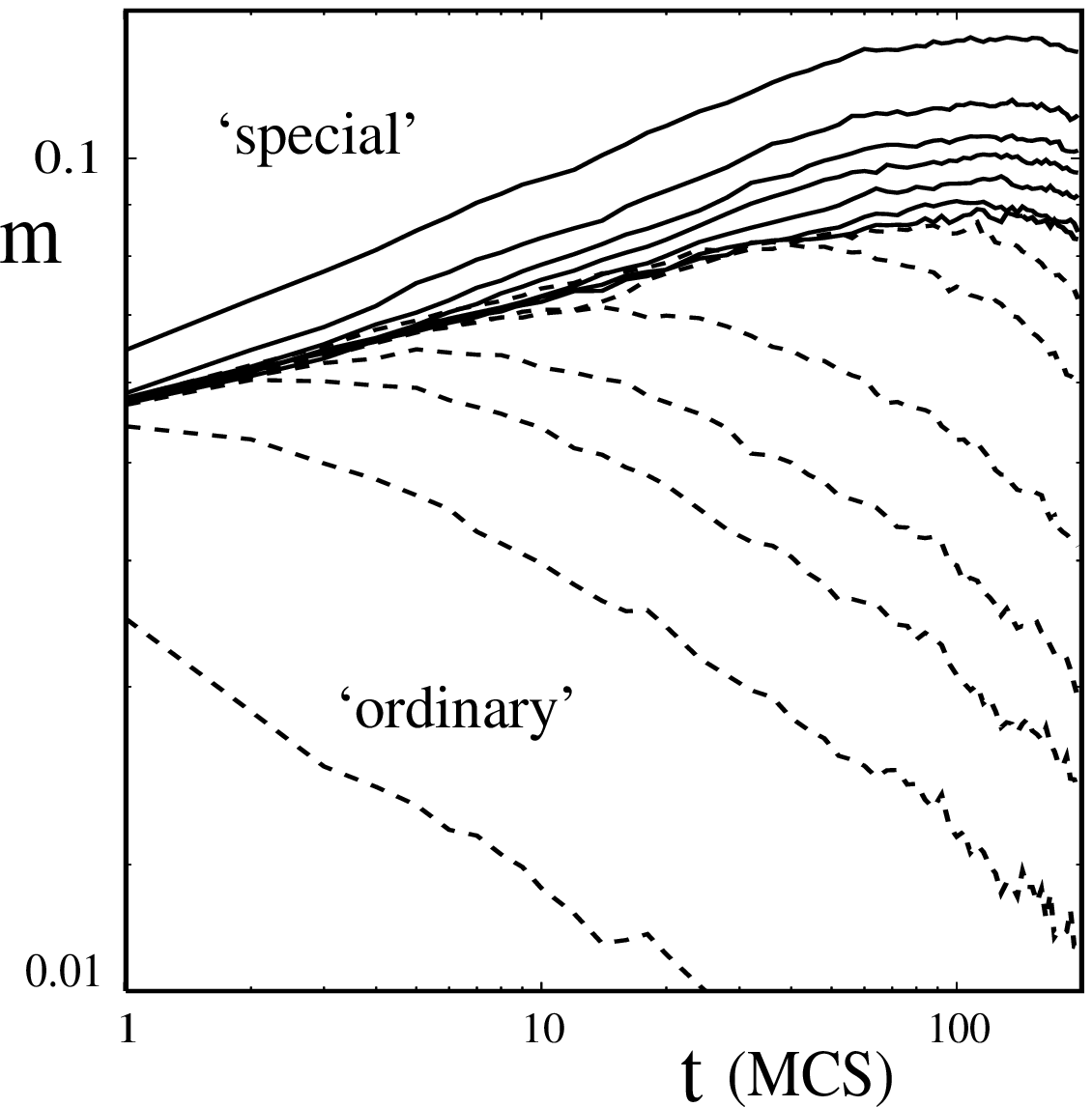}
\vspace*{-5.7cm}\\
\hspace*{9cm}
\noindent
\parbox{6cm}{{\small {\bf Fig 1.:} Monte-Carlo results from $20^2\times 40$
lattice
for the order parameter
for special (solid curves)
and ordinary (dashed curves)
transition. The distances from the surface are
$z=0,\,1,\,2,\,3,\,5,\,10,\,20$.
The behavior at the surface ($z=0$) is given by the uppermost
curve in the case of special and by the lowest curve
in the case of ordinary transition.}}
\\[1.0cm]

Results for the order parameter for special and ordinary
transition are shown in Fig.\,1 for different values of $z$.
As expected from the above discussion, all curves for points off
the surface start with bulk USTB. Close to the surface the profiles
show a crossover to the respective surface USTB.
Far off the surface this crossover does not occur
within the period of initial increase. For later times all curves
show the `linear' bulk behavior \cite{own,own2};
the nonlinear decay could only be seen in larger systems\cite{monte}.
The values for $\theta_1$, determined from our data,
are displayed in Table 1. Taking into account the errors, they
are consistent with the
expectation from (\ref{scalrel1}).
The situation is very similar for the autocorrelation function.
The results for $\lambda_1$ are also displayed in the Table.

To conclude, we have
investigated the critical dynamics, especially the
short-time behavior, of the $n$-vector model A
near `ordinary' and `special' surfaces.
By general scaling analysis
combined with a short-distance expansion, we have related
the surface short-time exponents to
other known exponents. The corresponding scaling relations are
given in
Eqs.\,(\ref{scalrel1}) and (\ref{scalrel2}).
In a one-loop calculation
these scaling relation were verified, and numerical
estimates for the surface
exponents were determined by Monte
Carlo simulation. We think that with the help of universal
short-time behavior, along the lines
suggested recently by Li et al. \cite{liea} for the bulk,
it should also be possible to determine {\it static}
surface exponents more precisely.
Further, we believe that short-time dynamics near surfaces
could also be an attractive field for experimental studies.\\
{\small {\it Acknowledgements}: We should like to thank H. W. Diehl
and F. Wichmann for useful suggestions and discussions.
This work was supported in part by the
Deutsche Forschungsgemeinschaft through
Sonderforschungsbereich 237.}


\begin{thebibliography} {99}
\bibitem{foo1} The short-time exponent $\theta$ was denoted by
$\theta'$
in Ref.\,\cite{jans}. But since the original $\theta$ of \cite{jans}
does not appear in the present work, we have dropped the prime.
Further, in order to avoid confusion with the spatial
coordinate $z$, we throughout call the dynamic equilibrium
exponent $\zeta$.
\bibitem{foo2}
For a more rigorous treatment we refer to Ref.\,\cite{jans2}.
\bibitem{foox} These results hold for both surface transitions discussed
here. For more details concerning the definition of static surface
exponents and scaling relations among them we refer to Ref.\,\cite{diehl}.
\bibitem{foo3} Since the `special' point is an unstable fixed-point of
the renormalization-group flow,
one should observe for all surface couplings $J_1$ lower than
the multicritical value the `ordinary' behavior provided the
system is large enough.
\end{thebibliography}
\end{document}